\documentclass[twocolumn,showpacs,preprintnumbers,amsmath,amssymb]{revtex4}

\usepackage{epsfig}
\usepackage{graphics}
\usepackage{graphicx}
\usepackage{dcolumn}
\usepackage{bm}

\begin{document}

\preprint{APS/123-QED}
\title{Intermittency transition to generalized synchronization in coupled time-delay
systems}

\author{D.~V.~Senthilkumar$^1$}
\email{skumar@cnld.bdu.ac.in}
\author{M.~Lakshmanan$^1$}%
 \email{lakshman@cnld.bdu.ac.in}

\affiliation{%
$^1$Centre for Nonlinear Dynamics,Department of Physics,
Bharathidasan University, Tiruchirapalli - 620 024, India\\
}%

\date{\today}

\begin{abstract}

In this paper, we report the  nature of transition to generalized
synchronization (GS) in a system of two coupled scalar piecewise linear
time-delay systems using the auxiliary system approach. We demonstrate that the
transition to GS occurs via on-off intermittency route and also it exhibits
characteristically distinct behaviors for different coupling configurations.  In
particular, the intermittency transition occurs in a rather broad range of
coupling strength for error feedback coupling configuration and in a narrow
range of coupling strength for direct feedback coupling configuration. It is
also shown that the intermittent dynamics displays periodic bursts of period
equal to the delay time of the response system in the former case, while they
occur in random time intervals of finite duration in the latter case.  The
robustness of these transitions with system parameters and delay times has also
been studied for both linear and nonlinear coupling configurations. The results
are corroborated  analytically by suitable stability conditions for
asymptotically stable synchronized states and numerically by the probability of
synchronization and by the transition of \emph{sub}Lyapunov exponents of the
coupled time-delay systems. We have also indicated the reason behind these
distinct transitions by referring to unstable periodic orbit theory of
intermittency synchronization in low-dimensional systems.

\end{abstract}

\maketitle

\section{\label{sec:level1}Introduction}

Synchronization of interacting chaotic oscillators is one of the most
interesting nonlinear phenomenon and is an inherent part of many natural
systems ~(cf.\cite{asp2001,sbjk2002}). The concept of synchronization is
receiving a central importance in recent research in nonlinear dynamics due to
its potential applications in diverse areas of science and technology.  Since
the identification of chaotic synchronization \cite{hfty1983,hfty198370,
lmptlc1990}, several works have appeared in identifying and demonstrating
basic kinds of synchronization both theoretically and experimentally 
\cite{asp2001,sbjk2002,hfty1983,hfty198370,lmptlc1990}.  There are also
attempts to find a unifying framework for defining the overall class of chaotic
synchronizations \cite{rblk2000,sblmp2001,aehaak2004}.

One of the interesting synchronization behaviors of unidirectionally coupled
chaotic systems is the generalized synchronization (GS), which was conceptually
introduced in Ref.~\cite{nfrmms1995}. Generalized synchronization is observed in
coupled nonidentical systems, where there exists some functional relation
between the  drive $X(t)$ and the response $Y(t)$ systems, that is,
$Y(t)=F(X(t))$.  With GS, all the response systems coupled to the drive lose
their intrinsic chaoticity (sensitivity to initial conditions) under the same
driving and follow the same trajectory. Hence the presence of GS can be detected
using the so called auxiliary system approach \cite{hdianfr1996}, where an
additional system (auxiliary system) identical to  the response system is
coupled to the drive in similar fashion. Auxiliary system approach is
particularly appealing since it can be implemented directly in an experiment
and, in addition, this method allows one to utilize analytical approaches for
studying GS.  However, one  has to be aware that if there are multiple basins of
attraction for the coupled drive-response system, then the auxiliary system
approach can  fail.

Generalized synchronization (GS) has been well studied and
understood in systems with few degrees of freedom and for discrete maps
~\cite{nfrmms1995,hdianfr1996,rb1998,lkup1996,kp1996,brheo1997,nfrctl2001}. 
The concept of GS has  also been extended to  spatially extended
chaotic systems such as coupled Ginzburg-Landau equations~\cite{aehaakpre2005}. 
Recently, the terminology intermittent generalized synchronization (IGS) 
\cite{aehaak2005} was introduced in diffusively coupled R\"ossler systems in
analogy with  intermittent lag synchronization (ILS) \cite{sbdlv2000,dlvsb2001}
and intermittent phase synchronization (IPS) \cite{apgo1997,apmz1997,kjlyk1998},
and also  experimentally  in coupled Chua's circuit.  Very recently, it has been
shown \cite{lzycl2005} that the transition to intermittent chaotic
synchronization (in the case of complete synchronization) is characteristically
distinct for geometrically different chaotic attractors.  In particular, it has
been shown that for phase coherent chaotic attractors (R\"ossler attractor) the
transition occurs immediately as soon as the coupling strength is increased from
zero and for non-phase-coherent attractors (Lorenz attractor),  the transition
occurs slowly as the coupling is increased from zero.

Time-delay systems form an important class of dynamical systems and recently
they are receiving central importance in investigating the phenomenon of chaotic
synchronizations in view of their infinite dimensional nature and feasibility of
experimental realization~\cite{mzxw2003,emskas2005,dvskmlpre2006,dvskml2005}. 
While the concept of GS has been well established in low dimensional systems, it
has not yet been studied in detail in coupled time-delay systems and  only very
few recent studies have been dealt with GS in time-delay
systems~\cite{mzxw2003,emskas2005}.  In particular, the mechanism of onset of GS
in coupled time-delay systems and its characteristic properties have not yet
been clearly understood and require urgent attention.

In this paper, we investigate the characteristic properties of nature of 
onset of GS from asynchronous state in unidirectionally coupled piece-wise
linear time-delay systems exhibiting highly non-phase coherent hyperchaotic
attractors~\cite{dvskmlpre2006}. We find that the onset of GS is preceded by
on-off intermittency mechanism from the desynchronized state. We have also
identified that the intermittency transition to GS exhibits characteristically
distinct behaviors for different coupling schemes.  In particular, the
intermittency transition occurs in a broad range of coupling strength for error
feedback coupling configurations and in a narrow range of coupling strength for
direct feedback  coupling configurations, beyond certain threshold value of
coupling strength. In addition, the intermittent dynamics is characterized by
periodic bursts away from the temporal synchronized state with period equal to
the delay time of the response system in the case of broad range intermittency
transition whereas it is characterized by random time intervals in the case of
narrow range intermittency transition.  We have also confirmed these dynamical
behaviours in both linear and nonlinear coupling configurations. We have
analyzed these transitions analytically using Krasvoskii-Lyapunov functional
approach and numerically by the probability of synchronization and by the
\emph{sub}Lyapunov exponents.  We have also addressed the reason behind these
transitions using periodic orbit theory. The robustness of these transitions
with the system parameters in both the linear and nonlinear, error feedback and
direct feedback coupling configurations, are also studied.

\begin{figure}
\centering
\includegraphics[width=0.5\columnwidth]{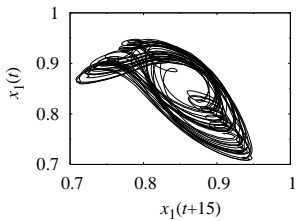}
\caption{\label{chaos} The hyperchaotic attractor of the system~(\ref{eq.onea})
 for the parameter values $a=1.0, b=1.2$ and $\tau=15.0$.}
\end{figure}

The plan of the paper is as follows. In Sec.~II, we will point out the existence
of broad range intermittency transition to GS for linear error feedback
coupling  proportional to ($x_1(t)-x_2(t)$) while in Sec.~III the existence of 
narrow range intermittency transition is shown for the linear direct feedback
coupling of the form $x_1(t)$, where $x_1(t)$ and $x_2(t)$ are the drive and
response signals, respectively (see below for details). In Sec.~IV we will
discuss the existence of broad range intermittency transition for nonlinear
error feedback coupling of the form $(f(x_1(t-\tau_2))-f(x_2(t-\tau_2)))$, where
$f(x)$ is an odd piece-wise linear function. The existence of narrow range
intermittency transition is discussed in Sec.~V for nonlinear direct feedback
coupling of the form $(f(x_1(t-\tau_2))$.  Finally in Sec.~VI, we will summarize
our results.

\section{Broad range (slow/delayed) intermittency transition to GS for linear
error feedback coupling of the form $(x_1(t)-x_2(t))$}

We consider the following first order delay differential equation introduced
by Lu and He~\cite{hlzh1996} and discussed in detail by Thangavel et
al.~\cite{ptkm1998},
\begin{eqnarray}
\dot{x}(t)&=&-ax(t)+bf(x(t-\tau)),
\label{eq.onea}
\end{eqnarray}
where $a$ and $b$ are parameters, $\tau$ is the time-delay and $f$ is an
odd piecewise linear function defined as
\begin{eqnarray}
f(x)=
\left\{
\begin{array}{cc}
0,&  x \leq -4/3  \\
            -1.5x-2,&  -4/3 < x \leq -0.8 \\
            x,&    -0.8 < x \leq 0.8 \\              
            -1.5x+2,&   0.8 < x \leq 4/3 \\
            0,&  x > 4/3. \\ 
         \end{array} \right.
\label{eqoneb}
\end{eqnarray}

\begin{figure}
\centering
\includegraphics[width=1.0\columnwidth]{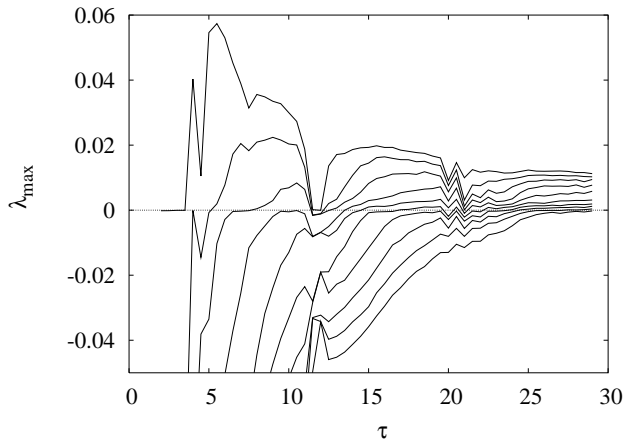}
\caption{\label{fig1} The first ten maximal Lyapunov exponents $\lambda_{max}$
 of the scalar time-delay system  for the parameter values
 $a=1.0,b_1=1.2, \tau\in(2,29)$ (which is the same as Eq.~(\ref{eq.onea}) with
 $b_1$ replaced by $b$) (\ref{eqonea}a).}
\end{figure}

Recently, we have reported ~\cite{dvskmlijbc}  that the system 
~(\ref{eq.onea}) exhibits hyperchaotic behavior for suitable parametric values.
For our present study, we find that for the choice of the parameters $a=1.0,
b=1.2$ and $\tau=15.0$ with the initial condition $x(t)=0.9, t\in(-5,0)$,
Eq.~(\ref{eq.onea}) exhibits hyperchaos.  The corresponding pseudoattractor is
shown in Fig.~\ref{chaos}. The hyperchaotic nature of Eq.~(\ref{eq.onea})
is confirmed by the existence of multiple positive Lyapunov exponents.   The
first ten maximal Lyapunov exponents for the above choice of parameters as a
function of delay time $\tau$ is shown in Fig.~\ref{fig1} (spectrum of Lyapunov
exponents in this paper are calculated using the procedure suggested by Farmer
\cite{jdf1982}).

To be specific, we first consider the following unidirectional, linearly
coupled systems with drive $x_1(t)$, response $x_2(t)$ and an auxiliary
$x_3(t)$,
\begin{subequations}
\begin{align}
\dot{x}_1(t)=&\,-ax_1(t)+b_{1}f(x_1(t-\tau_1)),  \\
\dot{x}_2(t)=&\,-ax_2(t)+b_{2}f(x_2(t-\tau_2))\nonumber \\
&\,+b_{3}(x_1(t)-x_2(t)),\\
\dot{x}_3(t)=&\,-ax_3(t)+b_{2}f(x_3(t-\tau_2))\nonumber \\
&\,+b_{3}(x_1(t)-x_3(t)),
\end{align}
\label{eqonea}
\end{subequations}
where $b_1, b_2$ and $b_3$ are constant parameters, and $\tau_1$ and $\tau_2$
are constant delay parameters. Note that when $b_1\ne b_2$ or $\tau_1 \ne
\tau_2$ or both, corresponding to parameter mismatches, we have
unidirectionally coupled nonidentical systems (Eq.~(\ref{eqonea}a) and
(\ref{eqonea}b)), while the auxiliary system is given by (\ref{eqonea}c) and
$f(x)$ is the odd piece-wise linear function (\ref{eqoneb}). The coupling in
(\ref{eqonea}b) may be also called a linear error feedback coupling.

For simplicity, we have chosen $b_1=b_2$ so that the time-delays $\tau_1$ and
$\tau_2$ alone introduce a simple form of parameter mismatch between the drive
$x_1(t)$ and the response $x_2(t)$.  We have chosen the values of parameters as
$a=1.0, b_1=b_2=1.2, \tau_1=20$ and $\tau_2=25$.  For this parametric choice,
in the absence of coupling, all the three systems (\ref{eqonea}) evolve
independently and exhibit hyperchaotic attractors, which is confirmed by the
existence of multiple positive Lyapunov exponents (Fig.~\ref{fig1}). The actual
value of the positive Lyapunov exponents for $\tau=20$ are
$0.00916$, $0.00759$, $0.00565$, $0.00283$ and $0.00073$ and for $\tau=25$ they are
$0.01234$, $0.01067$, $0.00886$, $0.00658$, $0.00386$, $0.00229$, $0.00123$
and $0.00033$. 
\subsection{Stability Condition}
With GS, as all the response systems under the same driving follow the same
trajectory, it is sufficient to identify the existence condition for
establishment of complete synchronization (CS) between the response $x_2(t)$
and the auxiliary $x_3(t)$ systems in order to achieve GS between the drive
$x_1(t)$ and the response $x_2(t)$ systems.

Now,  for CS to occur between the response $x_2(t)$ and the auxiliary $x_3(t)$
variables, we consider the time evolution of the difference system with the
state variable $\Delta=x_3(t)-x_2(t)$. It can be written for small values of
$\Delta$ as
\begin{eqnarray}
\dot{\Delta}=-(a+b_3)\Delta+b_2f^\prime(x_2(t-\tau_{2}))\Delta_{\tau_2},
\label{eq.difsys}
\end{eqnarray}
where
\begin{eqnarray}
f^\prime(x)=
\left\{
\begin{array}{cc}
            -1.5,&  -4/3 < x \leq -0.8 \\
            1,&    -0.8 < x \leq 0.8 \\              
            -1.5,&   0.8 < x \leq 4/3. \\
         \end{array} \right.
\label{eqonec}
\end{eqnarray}
\begin{figure}
\centering
\includegraphics[width=1.0\columnwidth]{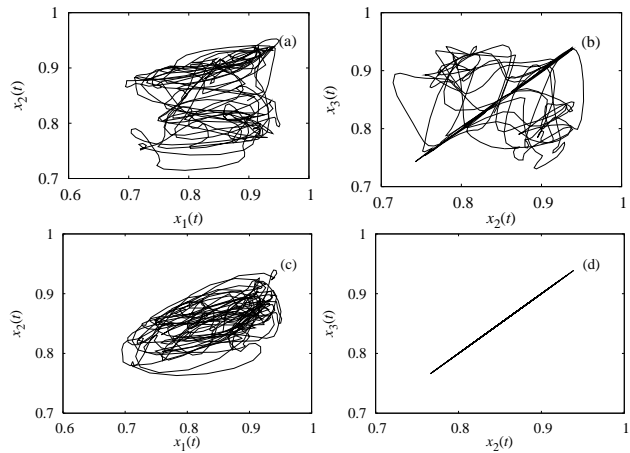}
\caption{\label{fig2} Dynamics in the phase space of the systems
(\ref{eqonea}). (a) and (b): Approximate GS and CS, respectively, for the value
of the coupling strength $b_3=0.4$. (c) and (d): Perfect GS and CS,
respectively, for the value of the coupling strength $b_3=0.9$.}
\end{figure}

The synchronization manifold, $x_2(t)=x_3(t)$, is locally attracting if the origin
of this equation is stable.  Following  Krasovskii-Lyapunov functional 
approach, we define a positive definite  Lyapunov functional of
the form \cite{nnk1963,kp1998,dvskml2005} (details of stability
analysis are given in appendix~\ref{a1})
\begin{eqnarray}
V(t)=\frac{1}{2}\Delta^2+\mu \int_{\tau_2}^0\Delta^2(t+\theta)d\theta,
\label{eq.lfunc}
\end{eqnarray}
where $\mu$ is an arbitrary positive parameter, $\mu>0$.  The solution of
Eq.~(\ref{eq.difsys}), namely $\Delta=0$, is stable if the derivative of the
functional along the trajectory of Eq.~(\ref{eq.difsys}) is negative.  This
negativity condition is satisfied if $b_3+a>\frac{b_2^2f^{\prime
2}(x_2(t-\tau_2))}{4\mu}+\mu$, from which it turns out that the sufficient
condition for asymptotic stability is
\begin{eqnarray}
a+b_3>\left|b_2f^{\prime}(x_2(t-\tau_2))\right|.
\label{eq.asystab}  
\end{eqnarray}
Now from the form of the piecewise linear function $f(x)$ given by
Eq.~(\ref{eqoneb}),
we have,
\begin{eqnarray}
\left|f^{\prime}(x_2(t-\tau_2))\right|=
\left\{
\begin{array}{cc}
1.5,&  0.8\leq|x_2|\leq\frac{4}{3}\\
1.0,&  |x_2|<0.8. \\
\end{array} \right.
\end{eqnarray}
Consequently the stability condition (\ref{eq.asystab}) becomes
$a+b_3>\left|1.5b_2\right|>\left|b_2\right|$.  Thus one can take
\begin{eqnarray}
a+b_3>\left|b_2\right|
\label{eq.ls} 
\end{eqnarray} 
as the less stringent or approximate stability condition (as the synchronization
dynamics of the coupled systems  (\ref{eqonea}) can occur even beyond the inner
region, $|x_2|<0.8$) for (\ref{eq.asystab}) to be valid, while 
\begin{eqnarray} 
a+b_3>\left|1.5b_2\right|
\label{eq.gc} 
\end{eqnarray} 
can be considered as the most general or stringent or exact stability condition
(as the full synchronization dynamics of the coupled systems  (\ref{eqonea})
lies within the outer region $0.8<|x_2|\le\frac{4}{3}$) specified by 
(\ref{eq.asystab}) for asymptotic stability of the synchronized state
$\Delta=0$.

\begin{figure}
\centering
\includegraphics[width=1.0\columnwidth]{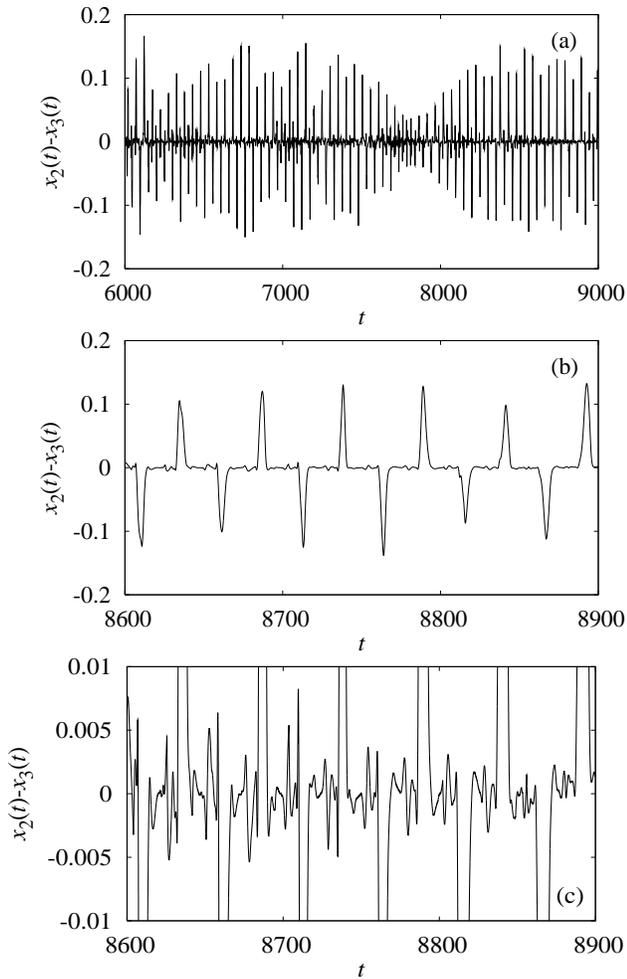}
\caption{\label{fig3} The intermittent dynamics of the response $x_2(t)$ and 
auxiliary $x_3(t)$ systems for the value of the coupling strength $b_3=0.4$. (a)
Time traces of the difference $x_2(t)-x_3(t)$ corresponding to Fig.~\ref{fig2}b,
(b) Enlarged in x scale to show bursts at periodic intervals when bursts of
larger amplitudes $\Delta>|0.01|$ are considered and (c) Enlarged in y scale to show random
bursts when bursts of smaller amplitudes $\Delta<|0.01|$ are considered.}
\end{figure}

\subsection{Approximate (Intermittent) Generalized Synchronization}
In order to understand the mechanism of transition to synchronized state, it
will be important to follow the dynamics from the parameter values at which the
less stringent condition is satisfied. Figure~\ref{fig2}a shows the approximate
GS (which may also be termed as intermittent generalized synchronization (IGS)
in analogy with intermittent lag synchronization (ILS)) between the drive 
$x_1(t)$ and the response $x_2(t)$ systems, whereas Fig.~\ref{fig2}b shows the
approximate CS between the response $x_2(t)$ and the auxiliary $x_3(t)$ systems 
for  the values of the parameters $a=1.0, b_1=b_2=1.2, \tau_1=20,  \tau_2=25$ 
and  $b_3=0.4$ satisfying the less stringent condition (\ref{eq.ls}). Perfect GS
and perfect CS are shown in Figs.~\ref{fig2}c and ~\ref{fig2}d respectively for
$b_3=0.9$ satisfying the general stability condition  (\ref{eq.gc}). Time traces
of the difference $x_2(t)-x_3(t)$ corresponding to approximate CS
(Fig.~\ref{fig2}b) are shown in Fig.~\ref{fig3}, which show periodic bursts with
period between two consecutive bursts approximately equal to the  time-delay of
the response system $t\approx 25$ when 'on' states of amplitude greater than
$|0.01|$ are considered. Fig.~\ref{fig3}b shows an enlarged (in x scale) 
part of Fig.~\ref{fig3}a to view the bursts at periodic intervals when bursts
of larger amplitudes ($\Delta>|0.01|$) are considered, while Fig.~\ref{fig3}c is
an enlarged (in y scale) version of  Fig.~\ref{fig3}b to show random bursts when
bursts of smaller amplitude $\Delta<|0.01|$ are considered. 

\begin{figure}
\centering
\includegraphics[width=1.0\columnwidth]{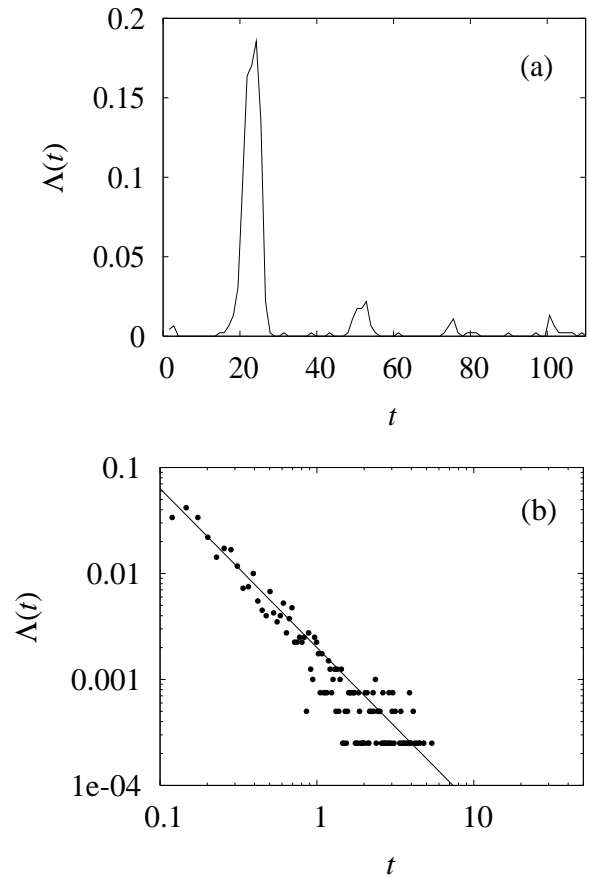}
\caption{\label{fig4} The statistical distribution of laminar phases
corresponding to the Fig.~\ref{fig3}. (a) For threshold value $\Delta=|0.1|$ and
(b) for the threshold value of $\Delta=|0.0001|$.}
\end{figure}

Usually the intermittent dynamics is characterized by the entrainment of the
dynamical variables in random time intervals of finite duration
\cite{npsmh1994,jfhnp1994}.  But from  Fig.~\ref{fig3}b, it is evident that the
intermittent dynamics displays periodic bursts from the synchronous state with
period approximately equal to the delay time of the response system, when
amplitudes of the state variable $|\Delta|=|x_3(t)-x_2(t)| > 0.01$ are
considered, for the values of the coupling strength at which the less stringent
stability condition (\ref{eq.ls}) is satisfied. The statistical features
associated with the intermittent  dynamics is analyzed by calculating the
distribution of laminar phases $\Lambda(t)$ with amplitude less than a threshold
value to $\Delta$.  A universal asymptotic power law distribution $\Lambda(t)
\propto t^{-\alpha}$ is observed for the threshold value $\Delta=|0.0001|$ with
the value of the exponent $\alpha=-1.5$ as shown in Fig.~\ref{fig4}b, which is
quite typical for on-off intermittency.  On the other hand the distribution of
laminar phases $\Lambda(t)$ for the value of the threshold value of
$|\Delta=0.1|$ shows a periodic structures (Fig.~\ref{fig4}a), where the peaks occur
approximately at $t=nT, n=1,2,...$, where $T\approx \tau_2$ is the period of the
lowest periodic orbit of the uncoupled system (\ref{eqonea}b). Note that $-3/2$
power law is observed for the intermittent dynamics shown in Fig.~\ref{fig3} for
laminar phases $\Lambda(t)$ with amplitude less than  $\Delta=|0.0001|$ (as an
illustrative example), which is also evident from Fig.~\ref{fig3}c, while
periodic bursts are observed for 'on' state of amplitude greater than $|0.01|$.
It is to be noted that such periodic bursts of period approximately equal to
the  time-delay of the response system for larger threshold value of $\Delta$
along with the on-off intermittency behavior for larger threshold value of
$\Delta$ has also been observed by Zhan \emph{et al} \cite{mzxw2003}  in
unidirectionally coupled Mackey-Glass systems , where the authors discussed
relation between two modes of synchronization, namely, CS and GS.

\begin{figure}
\centering
\includegraphics[width=1.0\columnwidth]{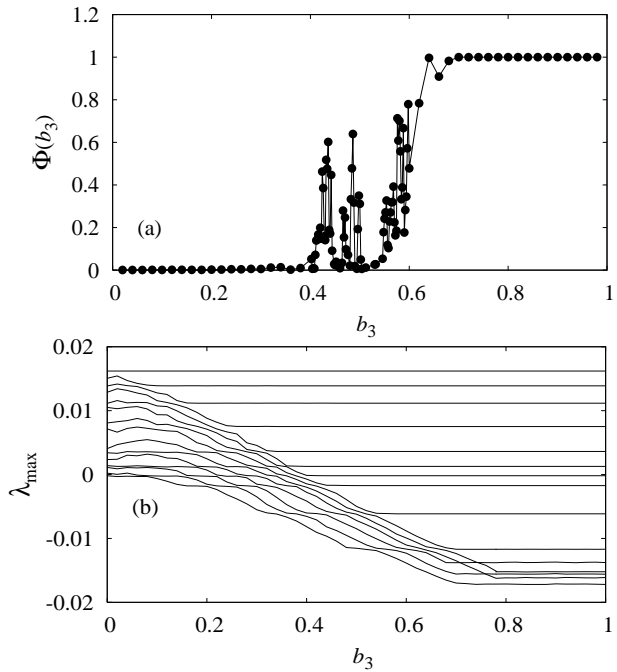}
\caption{\label{fig5} (a) The probability of synchronization $\Phi(b_3)$
between the response $x_2(t)$  and the auxiliary $x_3(t)$ systems and (b)
Largest Lyapunov exponents of the coupled drive $x_1(t)$ and response $x_2(t)$
systems (\ref{eqonea}a) and (\ref{eqonea}b).}
\end{figure}

\subsection{Characterization of IGS} Now we characterize the intermittency
transition to GS by using (i) the notion of the probability of synchronization
$\Phi(b_3)$ as a function of the coupling strength $b_3$~\cite{lzycl2005}, which
can be defined as the fraction of time during which $\left|x_2(t)-x_3(t)\right|<
\epsilon$ occurs, where $\epsilon$ is a small but arbitrary threshold, and (ii)
from the changes in the sign of \emph{sub}Lyapunov exponents (which are nothing
but the Lyapunov exponents of the subsystem) in the spectrum of Lyapunov
exponents of the coupled time-delay systems. Fig.~\ref{fig5}a shows the
probability of synchronization $\Phi(b_3)$ as a function of the coupling
strength $b_3$ calculated from the variables of the response $x_2(t)$ and the
auxiliary $x_3(t)$ systems for CS between them. For the range of
$b_3\in(0,0.39)$, there is absence of any entrainment between the systems
resulting in asynchronous behavior and the probability of synchronization
$\Phi(b_3)$ is practically zero in this region.  However, starting from the
value of $b_3=0.39$ and above, there appear  oscillations in the value of the
probability of synchronization $\Phi(b_3)$ between zero and some finite values
less than unity, exhibiting intermittency transition to GS in the range of
$b_3\in(0.4,0.62)$ for which the less stringent stability condition
(\ref{eq.ls}) is satisfied. Beyond $b_3=0.62$, $\Phi(b_3)$ attains unit value
indicating perfect GS. Note that the above intermittency transition occurs in a
rather wide range of the coupling strength (this can also be termed as slow or
delayed intermittency transition in analogy with the terminology used in
~\cite{lzycl2005}), which has also been confirmed from the transition of
successive largest \emph{sub}Lyapunov exponents in the corresponding range of
the coupling strength.

\begin{figure}
\centering
\includegraphics[width=1.0\columnwidth]{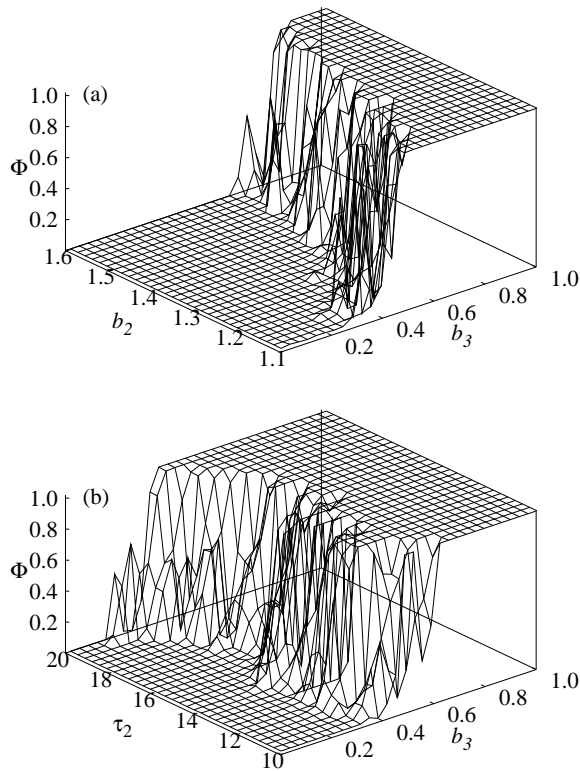}
\caption{\label{fig6} The probability of synchronization $\Phi(b_3)$ in
3-dimensional plots (a) as a function of the system parameter $b_2$ and the
coupling strength $b_3$ and (b) as a function of the coupling delay $\tau_2$
and the coupling strength $b_3$, exhibiting broad range intermittency transition
to GS for linear error feedback coupling.}
\end{figure}

The spectrum of the first fifteen largest Lyapunov exponents $\lambda_{max}$ of
the coupled  drive $x_1(t)$ and response $x_2(t)$ systems  are shown in
Fig.~\ref{fig5}b. From the general stability condition (\ref{eq.asystab}), it is
evident that for the chosen value of the parameter $a=1.0$, the less stringent
stability condition (\ref{eq.ls}) is satisfied for the values of coupling
strength $b_3>0.2$.   Correspondingly, the least positive \emph{sub}Lyapunov
exponent of the response system (\ref{eqonea}b) gradually becomes negative from
$b_3>0.2$. Subsequently, the remaining positive \emph{sub}Lyapunov exponents
gradually become negative and attain saturation in the range of
$b_3\in(0.2,0.8)$.  This is in accordance with the fact that the less stringent
stability condition is satisfied only in the corresponding range of coupling
strength $b_3$. This is a strong indication of the broad range intermittency
(IGS) transition to GS.  For $b_3>0.8$, the general stability condition
(\ref{eq.gc}) is satisfied, where one can observe perfect GS as is evidenced by
both the probability of synchronization approaching unit value and by the
negative saturation of \emph{sub}Lyapunov exponents calculated between the drive
and response systems. The inference is that the correlation between the
oscillations of the systems eventually becomes stronger with the strength of the
coupling, and this is indicated by the successive transition of
\emph{sub}Lyapunov exponents to negative values.

It is a well established fact that a chaotic attractor can be considered as a
pool of infinitely many unstable periodic orbits of all periods. 
Synchronization between two coupled systems is said to be asymptotically stable,
if all the unstable periodic orbits of the response system are stabilized in the
transverse direction of the synchroniation manifold. Consequently, all the
trajectories transverse to the synchronization manifold converge for suitable
values of coupling strength and this is reflected in the negative values of the
transverse Lyapunov exponents (\emph{sub}Lyapunov exponents) upon
synchronization~\cite{lzycl2005}. From our results, we find that the
\emph{sub}Lyapunov exponents gradually become negative in a broad range of
coupling strength $b_3$ after certain threshold value and this is in accordance
with the known results on gradual stabilization of unstable periodic orbits of
the response system in the complex synchronization manifold of low dimensional
systems~\cite{lzycl2005}. Unfortunately, methods for locating UPO's and
calculating their transverse Lyapunov exponents have not been well established
for time-delay systems and hence a quantitative proof for the gradual
stabilization of UPO's has not been given here.  However, the gradual
stabilization of UPO's along with their transverse Lyapunov exponent in the
range of intermittency transition have been reported for the case of coupled
R\"ossler and Lorenz systems in Ref.~\cite{lzycl2005}. It can then be inferred
from these studies that the broad range intermitteny transition in the case of
error feedback coupling configuration is due to the fact that the strength of
the coupling $b_3$ contributes only less significantly  to stabilize the UPO's
as the error $x_1(t)-x_2(t)$ gradually becomes smaller from the transition
regime after certain threshold value of the coupling strength. 

The robustness of the intermittency transition in a broad range of coupling
strength with the system parameter $b_2 \in (1.1,1.6)$  and with the coupling
delay $\tau_2 \in (10,20)$ has also been confirmed. Fig.~\ref{fig6}a shows the
3-dimensional plot of the probability of synchronization as a function of the
system parameter $b_2$ and the coupling strength $b_3$, while Fig.~\ref{fig6}b
shows the 3-dimensional plot of $\Phi(b_3)$ as a function of the coupling delay
$\tau_2$ and the coupling strength $b_3$. The above figures (Fig.~\ref{fig6})
clearly reveal the broad range intermittency transition to GS in the case of
linear error feedback coupling scheme.

\begin{figure}
\centering
\includegraphics[width=1.0\columnwidth]{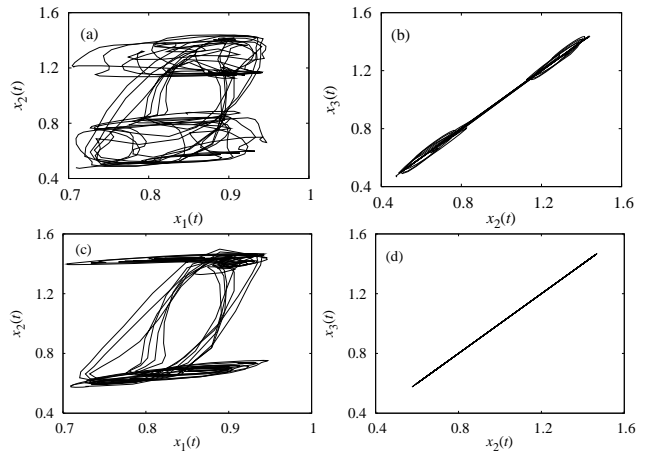}
\caption{\label{fig7} Dynamics in the phase space of the systems (\ref{lin2}):
(a) and (b) Approximate GS and CS, respectively, for the value of the coupling
strength $b_3=0.64$ and (c), (d) Perfect GS and CS, respectively, for the value of
the coupling strength $b_3=0.8$.}
\end{figure}

\section{\label{lin_narrow}Narrow range (immediate) intermittency transition to
 GS for linear direct feedback coupling of the form $x_1(t)$}
To illustrate the narrow range intermittency transition to GS, we consider the
unidirectional linear direct feedback coupling of the form
\begin{subequations}
\begin{eqnarray}
\dot{x}_1(t)&=&-ax_1(t)+b_{1}f(x_1(t-\tau_1)),  \\
\dot{x}_2(t)&=&-ax_2(t)+b_{2}f(x_2(t-\tau_2))+b_{3}x_1(t),\\
\dot{x}_3(t)&=&-ax_3(t)+b_{2}f(x_3(t-\tau_2))+b_{3}x_1(t),
\end{eqnarray}
\label{lin2}
\end{subequations}
where $f(x)$ is of the same odd piece-wise linear form as in 
Eq.~(\ref{eqoneb}). Assuming the same values of the parameters as before
and proceeding in the same way as in the previous case, one can obtain the
sufficient condition for asymptotically stable CS between the response $x_2(t)$
(\ref{lin2}b) and the auxiliary $x_3(t)$ (\ref{lin2}c) systems as 
\begin{eqnarray}
a>\left|b_2f^{\prime}(x_2(t-\tau_2))\right|.
\label{lin2stab}  
\end{eqnarray}

It is to be noted that the above stability condition holds good only for the
case when coupling is present, that is $b_3\ne0$.  When there is no coupling
($b_3=0$), by definition, there will be a desynchronized chaotic state. As soon
as the value of the coupling strength is increased from zero, the stability
condition (\ref{lin2stab}) always lead to synchronized state even for very
feeble values of $b_3$ for  parameters satisfying the stability condition, as
it  is independent of the coupling strength $b_3$. As the values of the
parameters satisfying the stability condition (\ref{lin2stab}) rapidly leads to
immediate transition to synchronized state as soon as the coupling is switched
on, it is difficult to identify the possible transitions to synchronized state. 
In addition, as the stability condition  is independent of the coupling strength
$b_3$, one is not  able to explore the dynamical transitions as a function of
coupling strength for the parameter values satisfying the stability condition
(\ref{lin2stab}). Hence we study the synchronization transition by choosing the
parameters violating the stability condition as $a=1.0, b_1=1.2, b_2=1.1,
\tau_1=\tau_2=20$ and varying the coupling strength $b_3$ in order to identify
the mechanism of synchronization transition. Here, in this case $b_1$ and $b_2$
alone introduce the parameter mismatch while $\tau_1=\tau_2$ (It may be added
that the qualitative nature of the dynamical transitions remain the same even
when the mismatch is either in time delays alone, that is  $\tau_1\neq\tau_2$, 
$b_1=b_2$ or in both the system parameters and time delays, $b_1\neq b_2$,
$\tau_1\neq \tau_2$, as confirmed below in the three dimensional plots of
Figs.~\ref{fig11}).

\begin{figure}
\centering
\includegraphics[width=1.0\columnwidth]{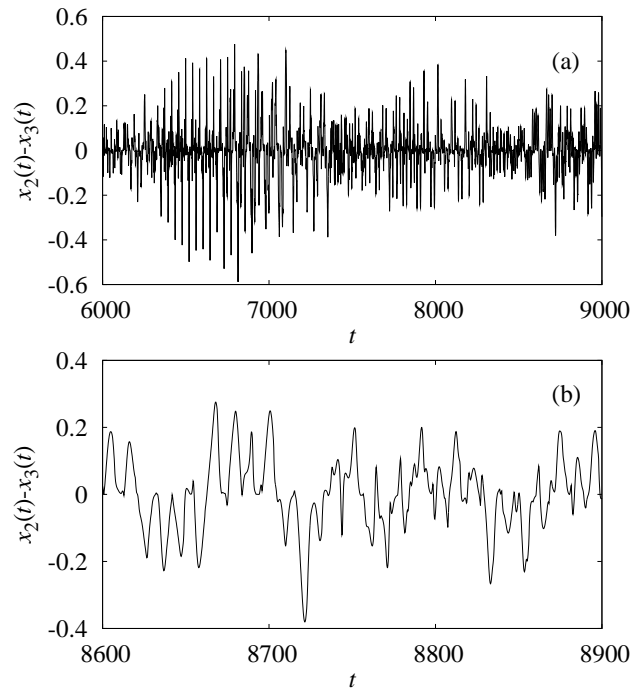}
\caption{\label{fig8} The intermittent dynamics of the response $x_2(t)$
(\ref{lin2}b) and  auxiliary $x_3(t)$ (\ref{lin2}c) systems for the value of the
coupling strength $b_3=0.64$. (a) and (b) Time traces of the difference
$x_2(t)-x_3(t)$ corresponding to Fig.~\ref{fig7}b.}
\end{figure}

\begin{figure}
\centering
\includegraphics[width=1.0\columnwidth]{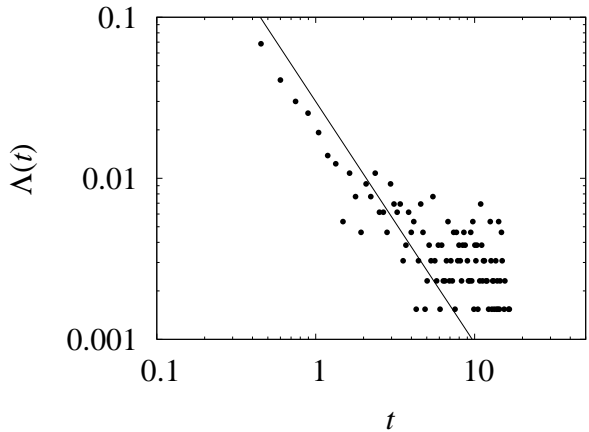}
\caption{\label{fig9} The statistical distribution of laminar phases for the
Fig.~\ref{fig8}.}
\end{figure}

As $b_3$ is varied from zero, transition from desynchronized state to
approximate GS occurs for $b_3>0.6$. Approximate GS (IGS) between the  drive
system $x_1(t)$ specified by Eq.~(\ref{lin2}a), and the response system 
$x_2(t)$ represented by  Eq.~(\ref{lin2}b), is shown in Fig.~\ref{fig7}a whereas
the approximate CS between the response $x_2(t)$ (Eq.~(\ref{lin2}b)) and the
auxiliary $x_3(t)$ (Eq.~(\ref{lin2}c)) systems is shown in  Fig.~\ref{fig7}b for
the value of the coupling strength $b_3=0.64$. Prefect GS and CS are shown in 
Figs.~\ref{fig7}c and \ref{fig7}d, respectively, for the value of the coupling
strength $b_3=0.8$. The intermittent dynamics at the transition regime
corresponding to the value of the coupling strength $b_3=0.64$ is shown in
Figs.~\ref{fig8}, in which Fig.~\ref{fig8}b shows the enlarged part of
Fig.~\ref{fig8}a. It is clear from this figure that the intermittent dynamics
displays intermittent bursts at random time intervals.  The statistical
distribution of the laminar phases again shows a universal asymptotic $-1.5$
power law behavior for the threshold value $\Delta=|0.0001|$, which is  typical
for on-off intermittency transitions, as shown in Fig.~\ref{fig9}.

\begin{figure}
\centering
\includegraphics[width=1.0\columnwidth]{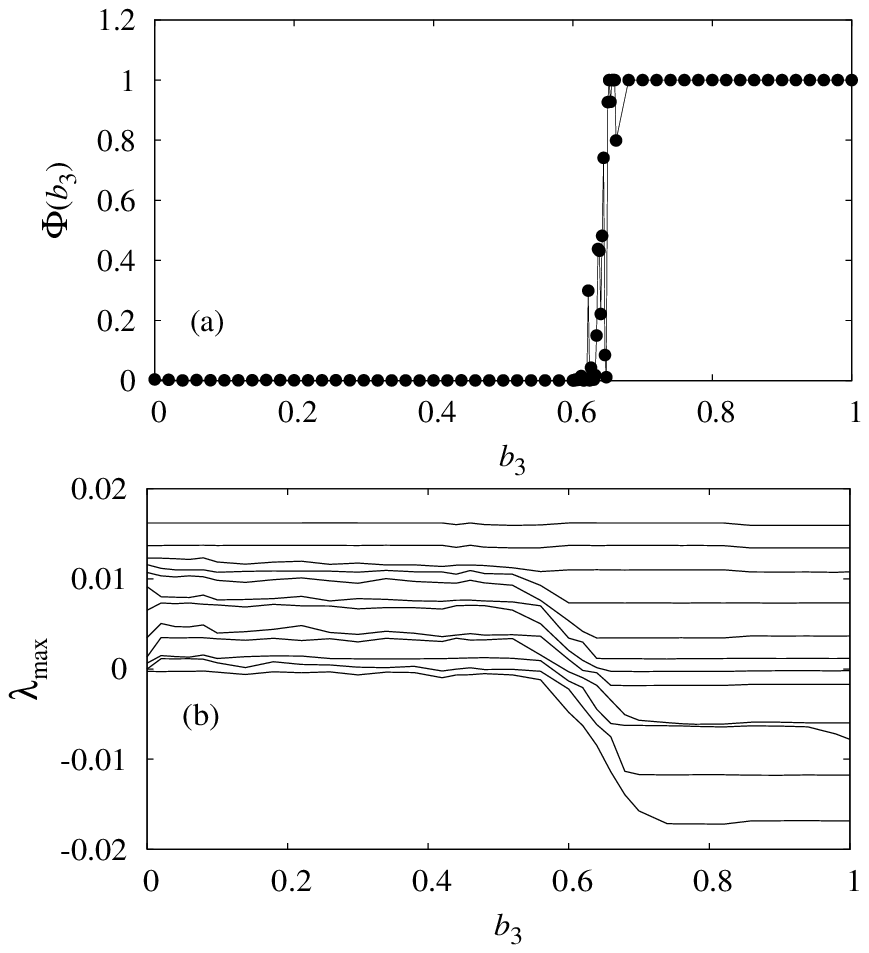}
\caption{\label{fig10} (a) The probability of synchronization $\Phi(b_3)$
between the response $x_2(t)$ (\ref{lin2}b) and the auxiliary  $x_3(t)$
(\ref{lin2}c) systems and (b) Largest Lyapunov exponents of the coupled drive
$x_1(t)$ and response $x_2(t)$ systems (\ref{lin2}a) and (\ref{lin2}b).}
\end{figure}

\begin{figure}
\centering
\includegraphics[width=1.0\columnwidth]{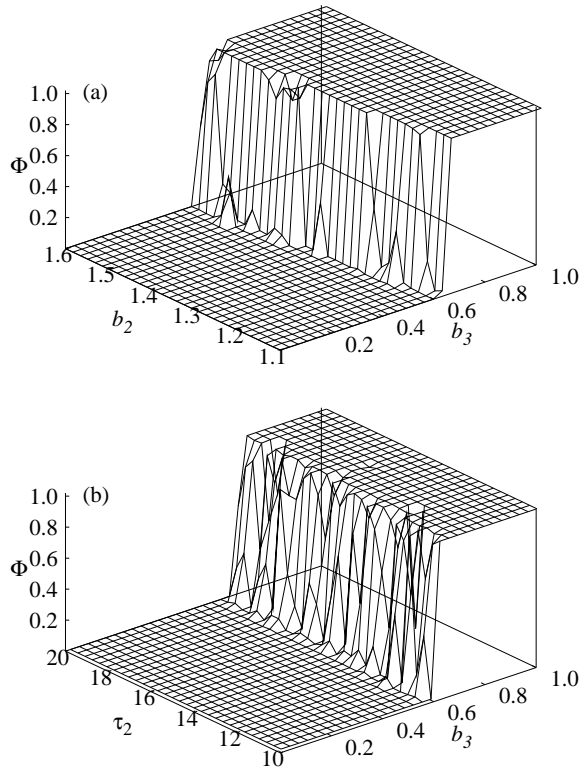}
\caption{\label{fig11} The probability of synchronization $\Phi(b_3)$ in
3-dimensional plots (a) as a function of the system parameter $b_2$ and the
coupling strength $b_3$ and (b) as a function of the coupling delay $\tau_2$
and the coupling strength $b_3$, exhibiting narrow range intermittency
transition to GS for linear direct feedback coupling}
\end{figure}

Now we characterize the intermittency transition to GS in the present case,
again by using the notion of the probability of synchronization $\Phi(b_3)$ and
from the changes in the sign of \emph{sub}Lyapunov exponents of the coupled
system. The probability of synchronization is shown in Fig.~\ref{fig10}a as a
function of the coupling strength, again calculated from the response $x_2(t)$
and  the auxiliary $x_3(t)$ systems, Eqs. (\ref{lin2}b) and (\ref{lin2}c),
respectively, which remains zero in the range of $b_3 \in (0,0.60)$ and
oscillates between its maximum and minimum values in a narrow range of $b_3 \in
(0.60,0.68)$ confirming the existence of approximate CS in the later range.
Above $b_3=0.68$ the probability of synchronization acquires its maximum value 
depicting perfect CS between the response $x_2(t)$  and the auxiliary $x_3(t)$ 
systems. Correspondingly there exists perfect GS between the drive $x_1(t)$  and
the response $x_2(t)$ systems. Fig.~\ref{fig10}b shows the first twelve  maximal
Lyapunov exponents of the coupled drive $x_1(t)$   and the response $x_2(t)$ 
systems. The least positive \emph{sub}Lyapunov exponent of the response system
starts to become negative from $b_3>0.60$. Subsequently, all the other positive
\emph{sub}Lyapunov exponents become negative and reach saturation in a rather
narrow range of $b_3 \in (0.60,0.68)$.  Thus the narrow range intermittency
(IGS) transition (this can also be termed as  immediate intermittency transition
in analogy with the terminology used in ~\cite{lzycl2005}) is confirmed from
both the probability of synchronization, calculated from the response and the
auxiliary systems, and negative saturation of \emph{sub}Lyapunov exponents,
calculated from the drive and the response systems.

As discussed in the previous section, the narrow range intermittency transition 
is in accordance with the stabilization of all the unstable periodic orbits of
the response system  in a narrow range as a function of the coupling strength
$b_3$ and this is reflected in the immediate transition of all the
\emph{sub}Lyapunov exponents (Fig.~\ref{fig10}b) to negative values.  It is also
to be noted that the narrow range intermitteny transition in the case of direct
feedback coupling configuration can be attributed to the fact that the strength
of the coupling $b_3$ contributes significantly proportional to the strength of
the signal $x_1(t)$ to stabilize all the UPO's  immediately at  the transition
regime after certain threshold value of the coupling strength as in the case of
low-dimensional systems~\cite{lzycl2005}.

The robustness of the intermittency transition in a narrow range of the coupling
strength $b_3$ for a range of values of the parameter $b_2 \in (1.1,1.6)$  and
the  delay $\tau_1=\tau_2 \in (10,20)$ is shown in Fig.~\ref{fig11}. The
3-dimensional plot of the probability of synchronization as a function of the
system parameter $b_2$ and the coupling strength $b_3$ is shown in  
Fig.~\ref{fig11}a, while Fig.~\ref{fig11}b shows the 3-dimensional plot of
$\Phi(b_3)$ as a function of the coupling delay $\tau_2$ and the coupling
strength $b_3$. 

\section{Broad range intermittency transition to GS for nonlinear error feedback
coupling of the form $(f(x_1(t-\tau_2))-f(x_2(t-\tau_2)))$}
Next we demonstrate the existence of the above types of distinct characteristic
transitions for nonlinear coupling configurations as well.  For this purpose,
we consider the  unidirectional nonlinear error feedback coupling of the form
\begin{subequations}
\begin{align}
\dot{x}_1(t)=&\,-ax_1(t)+b_{1}f(x_1(t-\tau_1)),  \\
\dot{x}_2(t)=&\,-ax_2(t)+b_{2}f(x_2(t-\tau_2)) \nonumber \\ 
&\, +b_{3}(f(x_1(t-\tau_2))-f(x_2(t-\tau_2))),\\
\dot{x}_3(t)=&\,-ax_3(t)+b_{2}f(x_3(t-\tau_2))\nonumber \\ 
&\, +b_{3}(f(x_1(t-\tau_2))-f(x_3(t-\tau_2))),
\end{align}
\label{nlin1}
\end{subequations}
where $f(x)$ is again of the same piece-wise linear form as in 
Eq.~(\ref{eqoneb}). The parameters are now fixed as $a=1.0,b_1=b_2=1.2,
\tau_1=20$ and the coupling delay $\tau_2=25$, where the delays alone form the
parameter mismatch between the drive and response systems in Eqs.~(\ref{nlin1}). Following
Krasvoskii-Lyapunov theory, for  complete synchronization so that the manifold
$\Delta=x_3(t)-x_2(t)$ between the response $x_2(t)$ and the auxiliary
$x_3(t)$ approaches zero, one can obtain the stability
condition  as
\begin{eqnarray}
a>\left|(b_2-b_3)f^{\prime}(x_2(t-\tau_2))\right|.
\label{nlinasystab}  
\end{eqnarray}
Consequently from the form of the piecewise linear function (\ref{eqoneb}),
the stability condition (\ref{nlinasystab}) becomes
$a>\left|1.5(b_2-b_3)\right|>\left|(b_2-b_3)\right|$. Thus one can take
\begin{eqnarray}
a>\left|b_2-b_3\right|
\label{nlin.ls}  
\end{eqnarray}
 as less stringent condition for (\ref{eq.asystab}) to be valid, while  
\begin{eqnarray}
a>\left|1.5b_2-1.5b_3\right|
\label{nlin.gc}  
\end{eqnarray}
can be considered as the most general condition specified by 
(\ref{nlinasystab}) for asymptotic stability of the synchronized state
$\Delta=x_2(t)-x_3(t)=0$. For the chosen values of the parameters, the less
stringent stability condition (\ref{nlin.ls}) is satisfied for the values of the
coupling strength $b_3\in(0.2,0.535)$ and the general stability condition
(\ref{nlin.gc}) is satisfied for $b_3>0.535$.

\begin{figure}
\centering
\includegraphics[width=1.0\columnwidth]{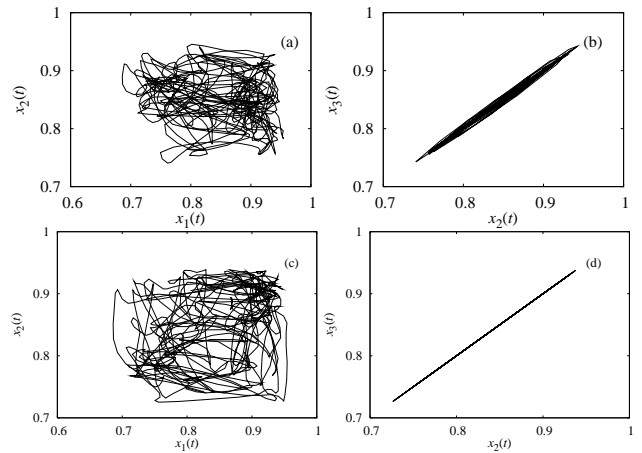}
\caption{\label{fig12} Dynamics in the phase space of the systems
(\ref{nlin1})). (a) and (b): Approximate GS and CS, respectively, for the value of the coupling
strength $b_3=0.37$. (c) and (d): Perfect GS and CS, respectively, for the value of the coupling
strength $b_3=0.8$.}
\end{figure}

\begin{figure}
\centering
\includegraphics[width=1.0\columnwidth]{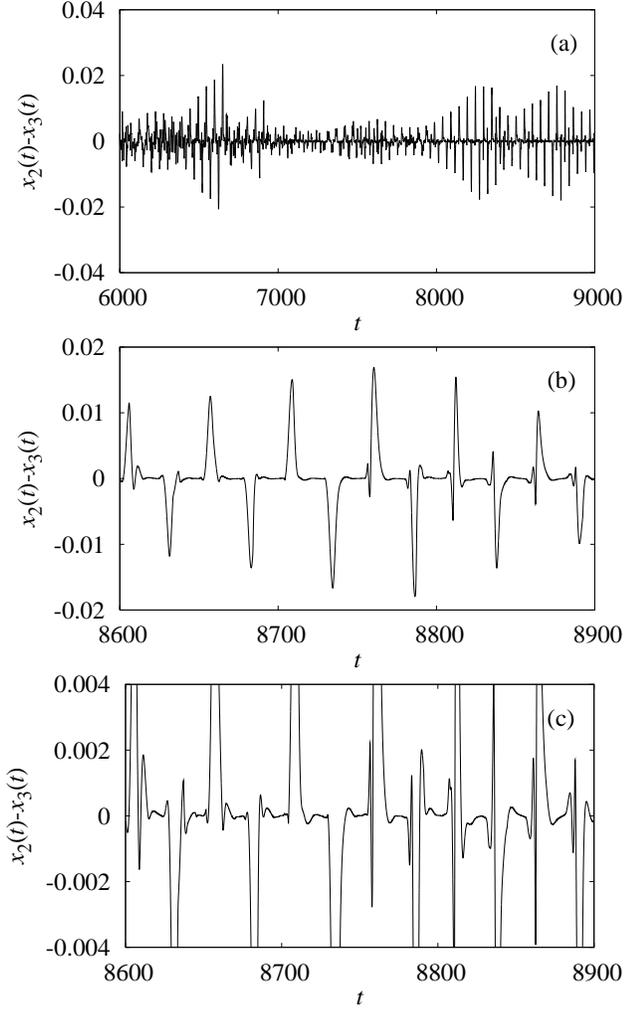}
\caption{\label{fig13} The intermittent dynamics of the response $x_2(t)$
(\ref{nlin1}b) and  auxiliary $x_3(t)$ (\ref{nlin1}c) systems for the the value
of the coupling strength $b_3=0.37$ for nonlinear error feedback coupling. (a)
Time traces of the difference $x_2(t)-x_3(t)$ corresponding to
Fig.~\ref{fig12}b, (b) Enlarged in x scale to show bursts at periodic intervals
when bursts of larger amplitudes $\Delta>|0.01|$ are considered and (c) Enlarged in y scale to
show random bursts when bursts of smaller amplitudes $\Delta<|0.01|$ are considered.}
\end{figure}

\begin{figure}
\centering
\includegraphics[width=1.0\columnwidth]{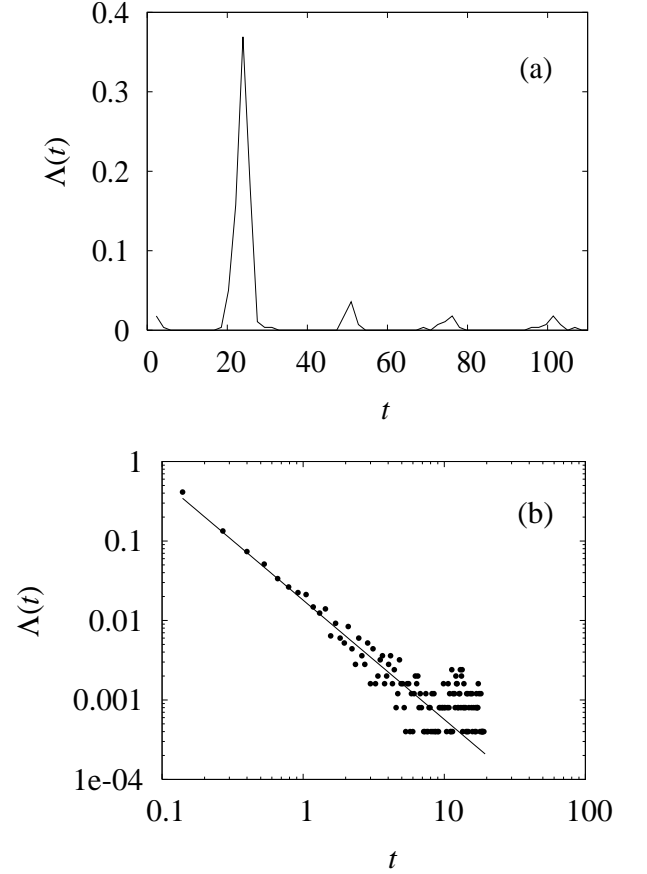}
\caption{\label{fig14} The statistical distribution of laminar phases
corresponding to  the
Fig.~\ref{fig13}. (a) For threshold value $\Delta=|0.1|$ and
(b) for the threshold value of $\Delta=|0.0001|$.}
\end{figure}

As the coupling strength is increased from zero, approximate GS occurs from
$b_3>0.2$. Figure \ref{fig12}a shows the approximate GS (IGS) between the drive
$x_1(t)$ (Eq.~(\ref{nlin1}a)) and the response $x_2(t)$ (Eq.~(\ref{nlin1}b))
systems for the value of the coupling strength $b_3=0.37$, while  the
approximate CS between the response $x_2(t)$ (Eq.~(\ref{nlin1}b)) and the
auxiliary $x_3(t)$  (Eq.~(\ref{nlin1}c))  systems  is shown in 
Fig.~(\ref{fig12}b). Perfect GS and perfect CS are shown in  Figs.~\ref{fig12}c
and \ref{fig12}d respectively for $b_3=0.8$. The intermittent dynamics exhibited
by the coupled systems at the transition regime is shown in Fig.~\ref{fig13},
which shows  bursts at the period approximately equal to the delay time of the
response system $x_2(t)$ for bursts of amplitude greater than $|0.01|$
(Fig.~\ref{fig13}b). The statistical distribution of the laminar phases away
from the intermittent bursts shows an asymptotic $-1.5$ power law behavior for
the threshold value $\Delta=|0.0001|$ (see Fig.~\ref{fig13}c), typical for
on-off intermittency, which is shown in Fig.~\ref{fig14}b. On the other hand for the
threshold value  $\Delta=|0.1|$ Fig.~\ref{fig14}a shows periodic structures
similar to Fig.~\ref{fig4}a with peaks occuring approximately at $t=nT,
n=1,2,...$, where $T\approx \tau_2$ is the period of the lowest periodic orbit
of the uncoupled system (\ref{nlin1}b).


\begin{figure}
\centering
\includegraphics[width=1.0\columnwidth]{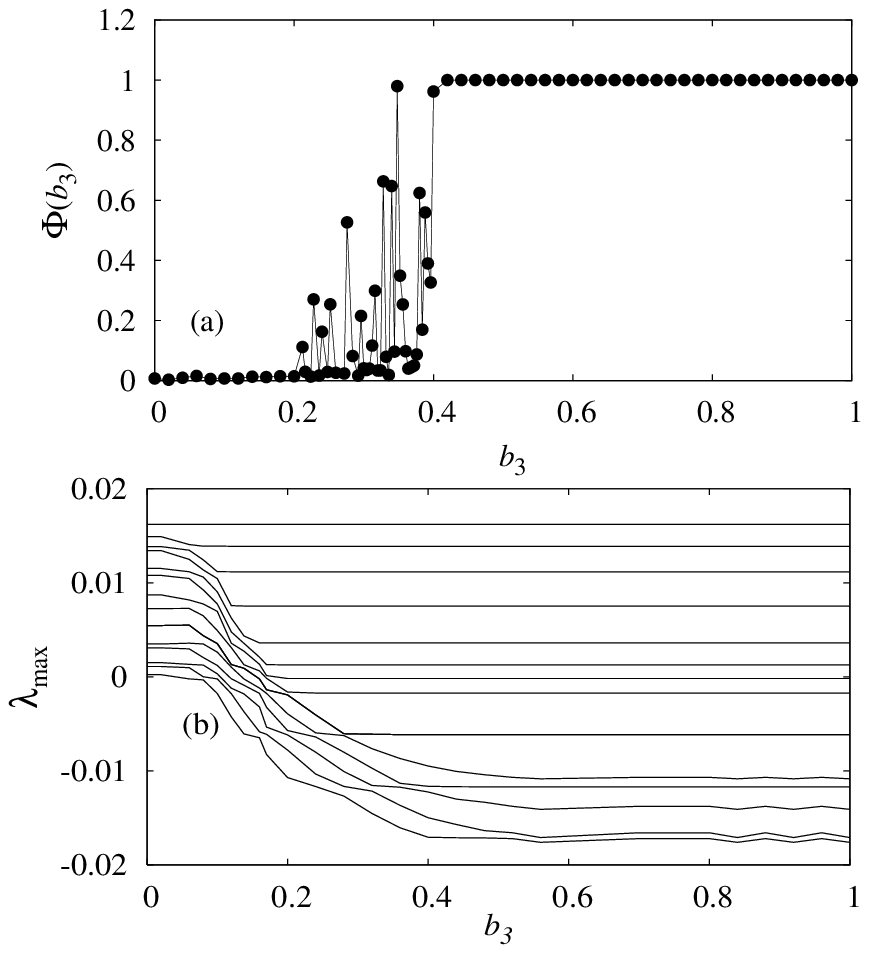}
\caption{\label{fig15} (a) The probability of synchronization $\Phi(b_3)$
between the response $x_2(t)$ (\ref{nlin1}b) and the auxiliary  $x_3(t)$
(\ref{nlin1}c) systems and (b) Largest Lyapunov exponents of the coupled drive
$x_1(t)$ and response $x_2(t)$ systems (\ref{nlin1}a) and (\ref{nlin1}b).}
\end{figure}

\begin{figure}
\centering
\includegraphics[width=1.0\columnwidth]{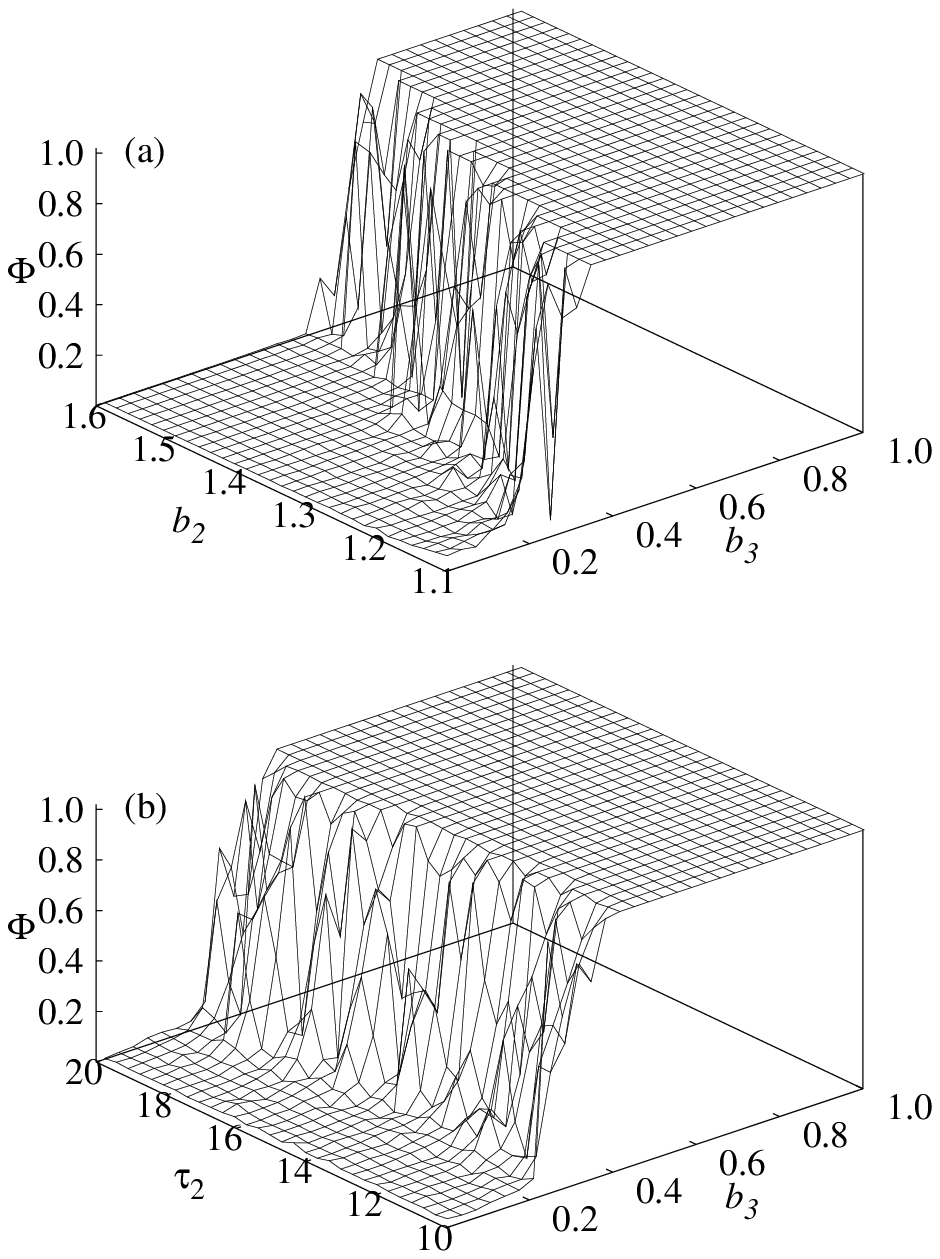}
\caption{\label{fig16} The probability of synchronization $\Phi(b_3)$ in
3-dimensional plots (a) as a function of the system parameter $b_2$ and the
coupling strength $b_3$ and (b) as a function of the coupling delay $\tau_2$ and
the coupling strength $b_3$ for the case of nonlinear error feedback coupling
configuration given by Eq.~(\ref{nlin1}), showing broad range intermittency
transition.}
\end{figure}

Now, the intermittency transition is again characterized using the probability
of synchronization and the \emph{sub}Lyapunov exponents as in the previous
cases. Fig.~\ref{fig15}a shows the probability of synchronization  $\Phi(b_3)$,
the value of which remains zero in the range $b_3 \in (0,0.2)$ due to the fact
that there lacks any entrainment between the response $x_2(t)$ and the auxiliary
$x_3(t)$ systems, whereas  it fluctuates between the two extreme values in a
rather broad range of the coupling strength  $b_3 \in (0.2,0.42)$, depicting the
existence of intermittency transition in the corresponding range of  $b_3$.
Perfect CS exists for $b_3>0.42$ as evidenced from the maximum value of 
$\Phi(b_3)$.  Correspondingly there exists perfect GS between the drive $x_1(t)$
and the response $x_2(t)$ systems. Figure \ref{fig15}b shows the transition of
\emph{sub}Lyapunov exponents of the spectrum of Lyapunov exponents of the
coupled drive $x_1(t)$ (Eq.~(\ref{nlin1}a)) and the response $x_2(t)$
(Eq.~(\ref{nlin1}b)) systems. The \emph{sub}Lyapunov exponents become negative
in the range  $b_3 \in (0.2,0.42)$ confirming the broad range intermittency
(IGS) transition in a rather wide range of the coupling strength and this is
again due to the gradual stabilization of the unstable periodic orbits of the
response systems because of the less significant contribution of the coupling
strength $b_3$ as the error becomes gradually smaller from the transition regime
beyond certain threshold value of the coupling strength as discussed in Sec.~II.
The robustness of the intermittency transition with the system parameter $b_2$
and the coupling delay $\tau_2$ as a function of coupling strength $b_3$ is
shown as 3-dimensional plots in Figs.~\ref{fig16}. 

\section{Narrow range intermittency transition to GS for nonlinear direct
feedback coupling of the form $f(x_1(t-\tau_2))$}

Now we consider the unidirectional nonlinear coupling of the form
\begin{subequations}
\begin{align}
\dot{x}_1(t)=&\,-ax_1(t)+b_{1}f(x_1(t-\tau_1)),  \\
\dot{x}_2(t)=&\,-ax_2(t)+b_{2}f(x_2(t-\tau_2))\nonumber \\
&\,+b_{3}f(x_1(t-\tau_2)),\\
\dot{x}_3(t)=&\,-ax_3(t)+b_{2}f(x_3(t-\tau_2))\nonumber \\
&\,+b_{3}f(x_1(t-\tau_2)).
\end{align}
\label{nlin2}
\end{subequations}

Choosing the values of the parameters  as in the previous case and
following Krasvoskii-Lyapunov functional approach for the asymptotically stable
synchronized state $\Delta=x_3(t)-x_2(t)=0$, one can obtain
the sufficient condition for asymptotic stability for complete synchronization
of the response $x_2(t)$ and the auxiliary $x_3(t)$ systems as  
\begin{eqnarray}
a>\left|b_2f^{\prime}(x_2(t-\tau_2))\right|. 
\label{nlin2asystab}  
\end{eqnarray}

\begin{figure}
\centering
\includegraphics[width=1.0\columnwidth]{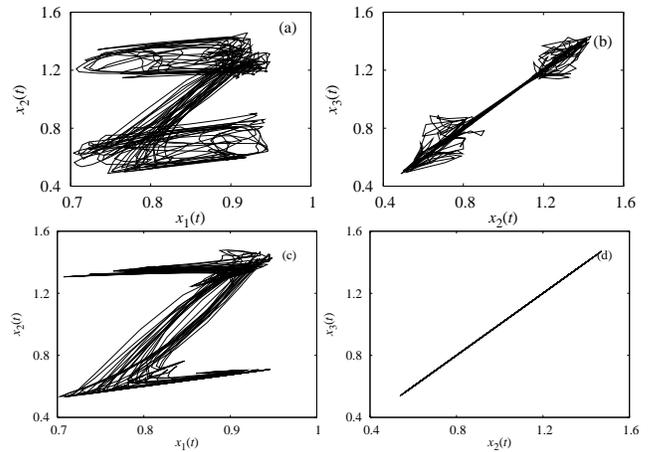}
\caption{\label{fig17} Dynamics in the phase space of the systems
(\ref{nlin2})). (a) and (b): Approximate GS and CS, respectively, for the value
of the coupling strength $b_3=0.78$. (c) and (d): Perfect GS and CS,
respectively, for the value of the coupling strength $b_3=0.9$.}
\end{figure}

The above stability condition rapidy leads to immediate transition to
synchronized state even for very feeble values of the coupling strength $b_3$
for the parameter values satisfying the stability condition 
(\ref{nlin2asystab}) as the stability condition is independent of 
$b_3$ as in the previous linear coupling case (Sec.~\ref{lin_narrow}). Hence
it is difficult to identify the possible dynamical transitions to synchronized
state as a function of the coupling strength $b_3$.  Hence we study the
synchronization transition as a function of the coupling strength $b_3$ by
choosing the parameters  violating the stability condition as $a=1.0,
b_1=1.2,b_2=1.1$ and $\tau_1=\tau_2=15$.

\begin{figure}
\centering
\includegraphics[width=1.0\columnwidth]{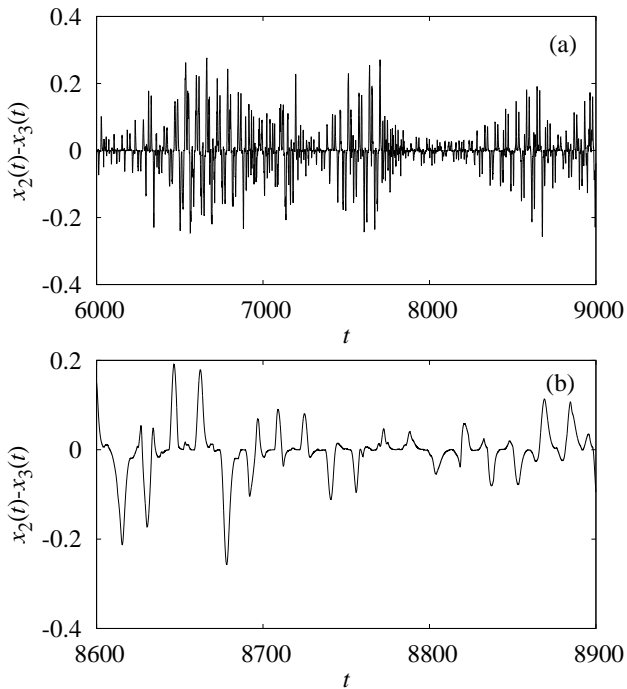}
\caption{\label{fig18} The intermittent dynamics of the response $x_2(t)$
(\ref{nlin2}b) and  auxiliary $x_3(t)$ (\ref{nlin2}c) systems for the the value
of the coupling strength $b_3=0.78$. (a) and (b) Time traces of the difference
$x_2(t)-x_3(t)$ corresponding to Fig.~\ref{fig17}b.}
\end{figure}

\begin{figure}
\centering
\includegraphics[width=1.0\columnwidth]{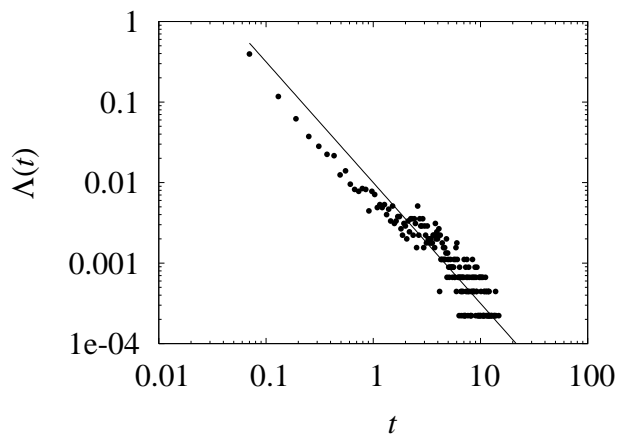}
\caption{\label{fig19} The statistical distribution of laminar phase for the
Fig.~\ref{fig18}.}
\end{figure}

As $b_3$ is varied from zero for the above values of the parameters, transition
from desynchronized state to approximate GS occurs for $b_3>0.74$. The
approximate GS (IGS) between the drive  $x_1(t)$  and the response $x_2(t)$ 
variables described by Eqs.(\ref{nlin2}a) and (\ref{nlin2}b) is shown in
Fig.~\ref{fig17}a, whereas Fig.~\ref{fig17}b shows the approximate CS between
the response $x_2(t)$  and the auxiliary $x_3(t)$  variables
(Eqs.~(\ref{nlin2}a) and (\ref{nlin2}c)) for  the value of the coupling strength
$b_3=0.78$. Perfect GS and perfect CS are shown in Figs.~\ref{fig17}c and
~\ref{fig17}d respectively for $b_3=0.9$. Time traces of the difference
$x_2(t)-x_3(t)$ corresponding to approximate CS (Fig.~\ref{fig17}b) is shown in
Fig.~\ref{fig18}, which shows intermittent dynamics with the entrainment of the
dynamical variables in random time intervals of finite duration. 
Fig.~\ref{fig18}b shows the enlarged picture of part of Fig.~\ref{fig18}a. The
statistical distribution of the laminar phases again shows a universal
asymptotic $-1.5$ power law behavior for the threshold value $\Delta=|0.0001|$, 
typical for on-off intermittency, as shown in Fig.~\ref{fig19}.

\begin{figure}
\centering
\includegraphics[width=1.0\columnwidth]{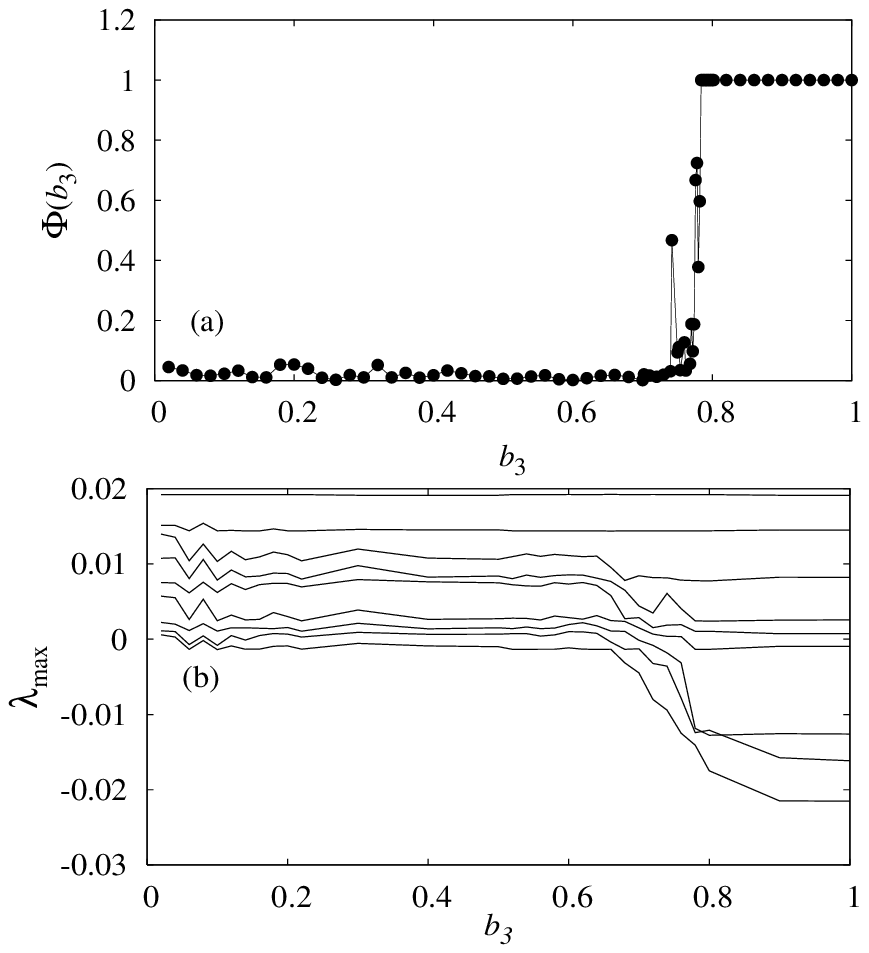}
\caption{\label{fig20} (a) The probability of synchronization $\Phi(b_3)$
between the response $x_2(t)$ (\ref{nlin2}b) and the auxiliary $x_3(t)$
(\ref{nlin2}c) systems and (b) Largest Lyapunov exponents of the coupled
drive $x_1(t)$ and response $x_2(t)$ systems (\ref{nlin2}a) and
(\ref{nlin2}b).}
\end{figure}

As in the previous cases, now we characterize the intermittency transition to GS
using  the notion of  probability of synchronization $\Phi(b_3)$ and  from the
changes in the sign of \emph{sub}Lyapunov exponents in the spectrum of Lyapunov
exponents of the coupled time-delay systems~(\ref{nlin2}). Fig.~\ref{fig20}a
shows the probability of synchronization $\Phi(b_3)$ as a function of the
coupling strength $b_3$ calculated from the response $x_2(t)$ and the auxiliary
$x_3(t)$ variables (Eqs.~(\ref{nlin2}b) and (\ref{nlin2}c)) for CS between
them.  In the range of $b_3\in(0,0.74)$, the probability of synchronization
remains approximately zero. Upon increasing the value of $b_3$, $\Phi(b_3)$
oscillates in the narrow range of  $b_3\in(0.74,0.78)$ depicting the existence
of intermittency transition. This  narrow range transition is also confirmed
from the transition of successive largest \emph{sub}Lyapunov exponents. The
spectrum of the first nine largest Lyapunov exponents $\lambda_{max}$ of the
coupled  drive $x_1(t)$  and response $x_2(t)$ variables (Eqs. (\ref{nlin2}a)
and  (\ref{nlin2}b)) is shown in Fig.~\ref{fig20}b. It is also evident from the
spectrum that the \emph{sub}Lyapunov exponents suddenly become negative in the
narrow range of $b_3\in(0.74,0.78)$, and then reach saturation values for
$b_3>0.78$. This confirms the narrow range intermittency (IGS) transition to GS.
This is also in accordance with the immediate stabilization of all the UPO's of
the response system as discussed in Sec.~(\ref{lin_narrow}).

\begin{figure}
\centering
\includegraphics[width=1.0\columnwidth]{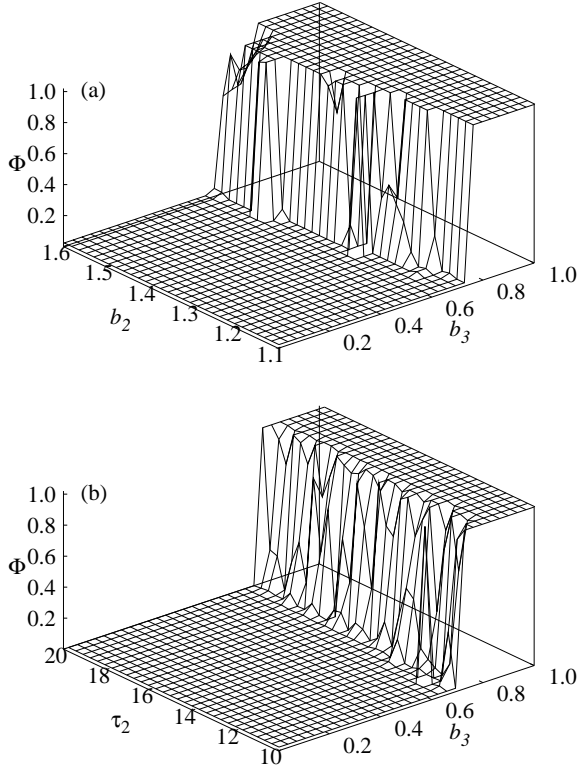}
\caption{\label{fig21} The probability of synchronization $\Phi(b_3)$ in
3-dimensional plots (a) as a function of the system parameter $b_2$ and the
coupling strength $b_3$ and (b) as a function of the coupling delay $\tau_2$
and the coupling strength $b_3$ for the case of nonlinear direct feedback
coupling configuration given by Eq.~(\ref{nlin2}) showing narrow range intermittency transition to GS.}
\end{figure}

The robustness of the transition with the system parameter $b_2$ and the delay
time $\tau_2$ as a function of coupling strength  $b_3$ is shown as
3-dimensional plots in Fig.~\ref{fig21}.

\section{Conclusion} 

In conclusion, we have studied intermittency transition to generalized
synchronization from desynchronized state in unidirectionally coupled scalar
piecewise linear time-delay systems for different coupling configurations using
the auxiliary  system approach. We have shown that the intermittency transition
to GS occurs in a  broad range of coupling strength  for both linear and
nonlinear error feedback coupling configurations whereas it occurs in a narrow
range of coupling strength for both the linear and nonlinear  direct feedback
coupling configuration. It has also been shown that the intermittent dynamics
displays periodic intermittent bursts of period equal to the delay time of the
response systems in the former cases and it takes place in  random time
intervals in the latter cases. The robustness of the intermittent dynamics with
the system parameters and the delay time  is also studied as a function of the
coupling strength for both  error feedback and direct feedback (linear and
nonlinear) coupling configurations. Universality of these intermittent behaviors
\cite{dvskml2006} (periodic and random) and their (broad and narrow range)
transitions are also confirmed for different forms of linear and nonlinear
coupling configurations in other time-delay systems such  as Mackey-Glass and
Ikeda systems. These distinct characteristic behaviors are analyzed using the
analytical stability condition for synchronized state, probability of
synchronization $\Phi(b_3)$ between the response and the auxiliary systems, and
by the changes in the sign of \emph{sub}Lyapunov exponents in the spectrum of
Lyapunov exponents of the drive and the response systems in both the linear and
nonlinearly coupled time-delay systems.  In spite of the fact that both the
probability of synchronization and the \emph{sub}Lyapunov exponents have been
calculated from different systems, we have found good agreement between them in
showing intermittency transition in all the cases. We hope this study will
contribute to the basic understanding of the nature of transition to  GS in
coupled time-delay systems and we are now investigating the experimental
verification of these findings in nonlinear electronic circuits.

\begin{acknowledgments}
The work of D. V. S and M. L has been supported by a Department of Science and
Technology, Government of India sponsored research project. The work of M. L is
supported by a DAE-BRNS Raja Ramanna Fellowship award, India.
\end{acknowledgments}
\appendix
\section{\label{a1} Stability condition}
To estimate a sufficient condition for the stability of the solution $\Delta=0$,
we require the derivative of the functional $V(t)$ along the trajectory
of Eq.~(\ref{eq.difsys}),
\begin{align}
\frac{dV}{dt}=-(a+b_3)\Delta^2+b_2f^{\prime}(x_2(t-\tau_2))\Delta
\Delta_{\tau_2}+\mu\Delta^2-\mu\Delta_{\tau_2}^2,
\end{align}
to be negative.  The above equation can be rewritten as
\begin{align}
\frac{dV}{dt}&=-\mu\Delta^2\Gamma(X,\mu),
\end{align}
where $X=\Delta_{\tau_2}/\Delta$, $\Gamma=\bigr[\bigr((a+b_3-\mu)/\mu\bigl)
-\bigr(b_2f^{\prime}(x_2(t-\tau_2))/\mu\bigl)X+X^2\bigl]$.
In order to show that $\frac{dV}{dt}<0$ for all $\Delta$ and $\Delta_{\tau_2}$
and so for all $X$, it is sufficient to show that $\Gamma_{min}>0$.
One can easily check that the absolute minimum of $\Gamma$ occurs at
$X=\frac{1}{2\mu}b_2f^{\prime}(x_2(t-\tau_2))$ with 
$\Gamma_{min}=\bigr[4\mu(a+b_3-\mu)-b_2^2f^{\prime}
(x_2(t-\tau_2))^2\bigl]/4\mu^2$.
Consequently, we have the condition for stability as
\begin{align}
a+b_3>\frac{b_2^2f^{\prime}(x_2(t-\tau_2))^2}{4\mu}+\mu = \Phi(\mu).
\label{eq.ineq}
\end{align}
Again $\Phi(\mu)$ as a function of $\mu$ for a given $f^{\prime}(x)$ has an
absolute minimum at $\mu=(|b_2f^{\prime}(x_2(t-\tau_2))|)/2$ with 
$\Phi_{min}=|b_2f^{\prime}(x_2(t-\tau_2))|$.  Since $\Phi\ge\Phi_{min}=
|b_2f^{\prime}(x_2(t-\tau_2))|$, from the inequality (\ref{eq.ineq}), it turns out that
the sufficient condition for asymptotic stability is
\begin{align}
a+b_3>|b_2f^{\prime}(x_2(t-\tau_2))|.
\label{a_eq.asystab}  
\end{align}

Now from the form of the piecewise linear function $f(x)$ given by
Eq.~\ref{eqoneb}, we have,

\begin{align}
|f^{\prime}(x_2(t-\tau_2))|=
\left\{
\begin{array}{cc}
1.5,& 0.8\leq|x_2|\leq\frac{4}{3}\\
1.0,& |x_2|<0.8 \\
\end{array} \right.
\end{align}
Note that the region $|x_2|>4/3$ is outside the dynamics of the present system
(see Eq.~(\ref{eqoneb})). Consequently the stability condition
(\ref{a_eq.asystab}) becomes $a+b_3>1.5|b_2|>|b_2|$.

Thus one can take $a+b_3>|b_2|$ as a less stringent condition for (\ref{a_eq.asystab}) to
be valid, while
\begin{align}
a>1.5|b_2|, 
\label{eq.four}
\end{align}
as the most general condition specified by (\ref{a_eq.asystab}) for asymptotic
stability of the synchronized state $\Delta=0$.


\end{document}